\documentclass[conference]{IEEEtran}
\IEEEoverridecommandlockouts
\usepackage{cite}
\usepackage{amsmath,amssymb,amsfonts}
\usepackage{algorithmic}
\usepackage{graphicx}
\usepackage{textcomp}
\usepackage{xcolor}
\usepackage{booktabs}
\usepackage{amsmath}
\usepackage{pifont}
\usepackage{amssymb}
\usepackage{makecell}
\usepackage{flafter}
\usepackage{longtable}
\usepackage{placeins}
\usepackage[hidelinks]{hyperref}
\usepackage{url}
\usepackage{microtype}
\usepackage{array}
\usepackage{multicol}

\def\BibTeX{{\rm B\kern-.05em{\sc i\kern-.025em b}\kern-.08em
    T\kern-.1667em\lower.7ex\hbox{E}\kern-.125emX}}
\begin{document}

\title{A Process Mining Software Comparison}
\author{\IEEEauthorblockN{Daniel Viner, Matthias Stierle, Martin Matzner}
\IEEEauthorblockA{{Institute of Information Systems, University of Erlangen-Nuremberg}\\
Nuremberg, Germany\\
\{daniel.viner, matthias.stierle, martin.matzner\}@fau.de}
}
\maketitle
\footnotesize Cite as: Daniel Viner, Matthias Stierle, and Martin Matzner. A process mining software comparison. In \textit{Proceedings of the ICPM Doctoral Consortium and Tool Demonstration Track 2020 co-located with the 2nd International Conference on Process Mining (ICPM 2020)}, volume 2703 of CEUR Workshop Proceedings,pages 19–22, 2020.

\begin{abstract}
www.processmining-software.com is a dedicated website for process mining software comparison and was developed to give practitioners and researchers an overview of commercial software available on the market. Based on literature review and experimental software testing, a set of criteria was developed in order to assess the tools’ functional capabilities in an objective manner. With our publicly accessible website, we intend to increase the transparency of software functionality. Being an academic endeavour, the non-commercial nature of the study ensures a less biased assessment as compared with reports from analyst firms.
\end{abstract}

\section{Introduction}
Starting in the late nineties as an academic research project, the discipline of process mining enjoys an increasing penetration in various industries over the past few years \cite{van2020}. Process mining helps organisations leverage event log data stored in databases or IT systems with the objective to discover, monitor and enhance processes \cite{van2016}.

The diversity of real-world applications is exemplified by the use of process mining software in banking, manufacturing, online gaming, healthcare, public service and many more industries \cite{hspi2020}. Use cases include compliance checks, continuous process improvement (CIP) and the assessment of robotic process automation (RPA) initiatives, to name a few. With the rise of use cases and continuous adoption of process mining in various industries, dozens of commercial tools have emerged on the process mining software market. The changing dynamics of the software market are marked by acquisitions, the increasing number of solutions and continuous releases of new features. Looking forward, the global process analytics market, which includes the discipline of process mining, is expected to grow at a rate of around 50\% annually from 2018 to 2023 to reach USD 1.42 billion by 2023 \cite{processanalytics.market}.
 
Depending on the scope and intended scalability, process mining initiatives may require high investments in terms of cost and stakeholder involvement, thus underscoring the danger of selecting the wrong software. Considering the academic context, process mining researchers are often not fully aware of practitioners’ needs and the developments in the software market. An overview of available tools and their capabilities is essential to address these issues. While several analyst firms such as Gartner \cite{gartner2019} published market studies that deliver an overview of the software landscape, we provide a more detailed analysis of process mining software with tangible criteria that examines functional capabilities. We conducted a non-commercial process mining software analysis and published the results on \url{www.processmining-software.com}. Besides serving as an independent software selection support for practitioners, the website also intends to help researchers understand the state-of-the-art in practice, allowing them to evaluate the usefulness of their work in regards to practical utility.

Based on literature review and experimental software testing, a set of criteria was derived in order to compare the features and functional capabilities of process mining software. This paper describes the underlying methodology and presents nine criteria categories with a brief description for each category and criterion.
\section{Methodology}
\subsection{Software Selection}
In order to ensure a comprehensive and representative listing, the most recent process mining-related reports of three analyst firms were taken as a basis to identify relevant software. Reports from Gartner \cite{gartner2019}, Everest Group \cite{everest2019,everest2020} and Forrester \cite{forrester2020} were analysed accordingly. Taking into consideration all software vendors stated in the commercial reports, a list of 34 potential tools was derived and further refined in three steps. First, three vendors not granting access to a demo environment were excluded from the study as we did not want to rely on information provided by the vendors. As the reports do not exclusively cover process mining software but also the software of related disciplines such as task mining or documentation, the respective tools were identified and excluded from the analysis in the second step, reducing the number of relevant vendors to 19. Third, three open-source tools were neglected. The ProM framework offers a comprehensible library of scientific techniques and algorithms but is geared towards academic scholars. PM4Py is an open-source Python library that currently does not provide a graphical user interface, making the solution difficult to use in the organisational context. Apromore was not tested either, however the tool will be considered in the second testing cycle due to the availability of a commercial license. Finally, 16 tools were tested, see Table \ref{tab:tools}.
\subsection{Evaluation Criteria}
In order to create a list of relevant criteria, a two-sided approach was followed. First, a literature review was undertaken to identify potential criteria from previous studies. The following terms were searched in the academic search engine sites WorldCat, SpringerLink and Google Scholar: “process mining software”, “process mining software comparison”, “process mining tools”, “process mining tool comparison” and “process mining criteria”. Second, the software was experimentally tested upfront to better understand what features and capabilities the vendors offer. Vendors were asked to grant access to all features in the demo environment to ensure all available features can be explored. The experimental approach also included the screening of all available knowledge bases and product documentations made accessible by the vendor. The derived criteria set was applied in three steps. In Phase 1, a test scenario was conducted for every tool using the same logs and files. In Phase 2, the results were compared with each other to identify inconsistent terminology and discrepancy in the level of detail. The final assessment was conducted in Phase 3. After testing, follow-up workshops were conducted with every vendor to clarify open questions and to get additional context for features. The exchange with the vendors also served as a quality gate for the correctness of the test results.
\subsection{Testing Setup}
The software testing was conducted primarily using event logs of Purchase-to-Pay (P2P) processes with their respective “happy path” reference models in BPMN format.
\section{Software Analysis}
\subsection{Analysed Software}
In the course of the study, 16 tools capable of mining event log files were analysed, see Table \ref{tab:tools}. The study was carried out in spring 2020.
\begin{table}[h!]
    \caption[Commercial process mining tools]{Commercial process mining software}
    \label{tab:tools}
    \centering
    \begin{tabular}{>{\raggedright\arraybackslash}m{3.75cm}>{\raggedright\arraybackslash}m{3.75cm}}
    \toprule
    \multicolumn{2}{c}{\textbf{Tool Name (Vendor)}}\\
    \midrule
    ABBYY Timeline (ABBYY) & MEHRWERK ProcessMining (Mehrwerk GmbH)\\
    \midrule
    ARIS Process Mining\newline (Software AG) & Minit (Minit j.s.a.)\\
    \midrule
    BusinessOptix (BusinessOptix) & myInvenio (myInvenio Srl)\\
    \midrule
    Celonis Process Mining\newline (Celonis SE) & PAFnow (Process Analytics \newline Factory GmbH)\\
    \midrule
    Disco (Fluxicon BV) & ProDiscovery\newline (Puzzle Data Co., Ltd.)\\
    \midrule
    EverFlow (EverFlow) & QPR ProcessAnalyzer\newline (QPR Software Plc)\\
    \midrule
    LANA Process Mining\newline (Lana Labs GmbH) & Signavio Process Intelligence (Signavio GmbH)\\
    \midrule
    Logpickr Process Explorer 360 (Logpickr) & UiPath Process Mining\newline (UiPath Inc.)\\
    \bottomrule
    \end{tabular}
\end{table}

\subsection{Website}
The website is mainly built on three layers. While the homepage (first layer) introduces the discipline of process mining, typical use cases and our criteria overview, the “Tools” page (second layer) lists brief profiles of all tools which are linked to the detailed tool profile pages (third layer). An introductory paragraph briefly describes the vendor and the strengths of its software. Eight criteria categories examine the availability and extent of tested functionality while one criteria category provides general information. The “Distinctive Focus and Features” section provides additional context by highlighting outstanding functionality. In order to offer users visual impressions of a tool, every profile is enriched with a “featured video” provided by the vendor and up to seven screenshots, of which five are defined and two undefined (proprietary). Also, any two selected tool profiles can be contrasted with each other through a side-by-side comparison.

\subsection{Software Criteria}
The software criteria derived from the literature review and experimental software testing represents the core of this study. The criteria were grouped into nine categories depicted in Tables \ref{tab:general_info} - \ref{tab:security} in the appendix.

Category \emph{General} gives a brief overview of the vendor and key aspects of the tool.
\emph{Data Management} examines functionalities and factors related to the extraction, transformation and loading (ETL) of process data into the process mining tool. 
The \emph{Process Discovery} category examines process graph capabilities and process analysis features such as benchmarking and rework analysis.
\emph{Conformance Checking} is a fundamental process mining feature to identify deviations between the actual “as is” process and an “a-priori” reference model. This category considers all relevant factors pertaining to conformance checking.
The \emph{Operational Support} criteria examine the availability of forward-looking capabilities to help users anticipate the outcome of running cases and facilitate decision making with the help of intelligent recommendations.
\emph{Views, Monitoring and Reporting} addresses the ability to monitor processes with the help of metrics and visualisations to support decision making. Additional criteria examine available languages and means of collaboration to share insights with other users.
While process enhancement functionality such as performance metrics in the process graph is partly covered in the aforementioned criteria categories, \emph{Advanced Enhancement Capabilities} investigates further capabilities that add a new perspective to the graph or the overall process. Lastly, \emph{Security \& Compliance} addresses role-based access control and the availability of audit logs.

\section{Contributions, Limitations and Outlook}
The study of 16 process mining solutions with commercial licenses showed that the maturity level of the investigated software is highly varying. While some vendors offer basic discovery functionality without conformance checking in some cases, other vendors offer more elaborate features such as process simulation, predictive analytics and decision rule mining. We observe a potential trend: The boundaries between mere process mining functionality and other disciplines such as process modelling (BPMN), business intelligence and Machine Learning become more and more blurred.

The software selection is based on software listed in commercial reports and hence reflects a non-exhaustive picture of the market. Further, open-source software was not analysed. It is important to note that the software listing represents only a snapshot of the tools’ capabilities and features in terms of information timeliness. Vendors are continuously improving their products and extend the functionalities with periodic releases.

A follow-up study could examine the perspective of organisations on the relevance of the suggested criteria. Interviews may be conducted with organisations interested in process mining as well as organisations with already implemented process mining software.

\section*{Acknowledgement}
The authors would like to express their gratitude to all vendors that participated in this study for their time and effort.

\bibliographystyle{IEEEtran}
\bibliography{references_new.bib}

\appendix

\begin{table}[htb]
    \caption[General Information]{General Information}
    \label{tab:general_info}
    \centering
    \begin{tabular}{>{\raggedright\arraybackslash}m{2cm}m{5.5cm}}
    \toprule
    \textbf{Criterion} & \textbf{Brief description}\\
    \midrule
   Company Size & 1-10, 11-50, 51-100, 101-250, 251-500,\newline 501-1000, 1001-5000, 5000+ employees\\
    \midrule
   Free Trial & “Immediate access” or “Upon request”\\
    \midrule
   Licenses & List of all available licenses types, e.g. Academic, Commercial\\
    \midrule
   Deployment & List of all available deployment options, e.g. On-Premises\\
    \midrule
   Embedded In & If applicable: Name of external system/platform that the software is embedded in, e.g. Qlik Sense\\
    \midrule
   Tested Version & Version/build number and month/year of testing\\
    \bottomrule
    \end{tabular}
\end{table}

\begin{table}[htb]
    \caption[Criteria Category “Data Management”]{Criteria Category “Data Management”}
    \label{tab:data}
    \centering
    \begin{tabular}{>{\raggedright\arraybackslash}m{2cm}m{5.5cm}}
    \toprule
    \textbf{Criterion} & \textbf{Brief description}\\
    \midrule
   Import File Types & Supported file types for event log upload, e.g. CSV, XES\\
    \midrule
   Database Connections & Available connectors to source data from databases, e.g. ODBC or JDBC drivers\\
       \midrule
   Adapters/ Connectors & Available connectors to source data from IT systems (e.g. ERP, CRM) or via APIs\\
       \midrule
   Integrated ETL Functionality & Yes/No - User can extract data from source system, perform $\geq$5 different transformation operations, and finally load data into the software\\
       \midrule
   Data Pseudo-\newline nymisation & Yes/No - Selected set of data can be pseudonymised, i.e. replaced with hash values\\
    \midrule
   Data Loading & Data Refresh \ding{51}\ding{55}, Scheduled Jobs \ding{51}\ding{55} \newline (User can append a new data set A to an existing data set B, i.e. incremental data loading; a time interval or specific dates for data extraction from a specified source can be configured)\\
    \midrule
   Character Encodings & Test of UTF-8 compatibility incl. special characters from various non-Latin languages; List of all additional supported character encodings\\
       \midrule
   Attribute Types & Case-level \ding{51}\ding{55}, Event-level \ding{51}\ding{55}\\
       \midrule
   Specify Business Hours & Working week \ding{51}\ding{55}, Multiple shifts/day \ding{51}\ding{55}, \newline Exclude days \ding{51}\ding{55}, Holiday calendar \ding{51}\ding{55}\\
       \midrule
   Define Event Order & List of manual means to order events in case of identical timestamps, e.g. by selected column that contains sorting information\\
       \midrule
   Start/End Timestamp & “1 timestamp”  or  “2 timestamps”\\
    \bottomrule
    \end{tabular}
\end{table}

\begin{table}[htb]
    \caption[Criteria Category “Process Discovery” (1/2)]{Criteria Category “Process Discovery” (1/2)}
    \label{tab:discovery}
    \centering
    \begin{tabular}{>{\raggedright\arraybackslash}m{2cm}m{5.5cm}}
    \toprule
    \textbf{Criterion} & \textbf{Brief description}\\
    \midrule
   As-Is Process Visualisation & List of all available visualisation types for the process graph, e.g. Directly-Follows Graph\\
    \midrule
   Export As-Is Process Graph & Available export formats for the process graph\\
    \midrule
   Performance Highlighting & Active time \ding{51}\ding{55}, Idle time \ding{51}\ding{55} \newline (Visual bottleneck highlighting of activities, i.e. active time, and transitions, i.e. idle time)\\
    \midrule
   Process Animation (Replay) & Adjust speed \ding{51}\ding{55}, Adjust timeframe \ding{51}\ding{55}, \newline Switch time mode \ding{51}\ding{55}, Zoom in case \ding{51}\ding{55} \newline (Animated replay of all process flows from a case perspective)\\
    \midrule
   Search \& Filter in Graph & Search \ding{51}\ding{55}, Filter \ding{51}\ding{55} \newline (User can search for any activity in the process graph; User can filter by activities/transitions (nodes/arcs) directly from the process graph)\\
    \midrule
   Graph Abstraction & Yes/No - Amount of displayed nodes and arcs in the process graph can be varied/adjusted\\
    \midrule
   Frequency Metrics & List of all available frequency-related metrics\\
    \midrule
   Time Metrics & List of all available performance-related metrics\\
    \midrule
   Additional Graph Metrics & Cost metrics \ding{51}\ding{55}, Custom metrics \ding{51}\ding{55}\\
    \bottomrule
    \end{tabular}
\end{table}

\begin{table}[ht!]
    \caption[Criteria Category “Process Discovery” (2/2)]{Criteria Category “Process Discovery” (2/2)}
    \label{tab:discovery2}
    \centering
    \begin{tabular}{>{\raggedright\arraybackslash}m{2cm}m{5.5cm}}
    \toprule
    \textbf{Criterion} & \textbf{Brief description}\\
    \midrule
   Process Benchmarking & Visual comparison \ding{51}\ding{55}, Metric comparison \ding{51}\ding{55} \newline (2 filtered sets of the same process can be compared with each other visually and metrically)\\
    \midrule
   Process Benchmarking (Different Logs) & Visual comparison \ding{51}\ding{55}, Metric comparison \ding{51}\ding{55} \newline (Processes of $\geq$2 different event logs can be compared with each other visually and metrically)\\
    \midrule
   Root Cause Analysis & Yes/No - The software delivers a list of root causes for selected or defined anomalies/symptoms\\
    \midrule
   Variant Breakdown by & List of metrics by which the variants can be classified/sorted\\
    \midrule
   Case and Activity List & Activity List \ding{51}\ding{55}, Case List \ding{51}\ding{55}, \newline Case List for Variants \ding{51}\ding{55}\\
    \midrule
   View Case Details & Yes/No - User can access a case view with respective case activities and metrics\\
    \midrule
   Rework Analysis & Yes/No - User can identify rework, i.e. loops and self-loops, through pre-configured dashboards or filtering\\
    \midrule
   Edge/Transition Details & List of all transitions \ding{51}\ding{55}, From-to activities \ding{51}\ding{55}\\
    \bottomrule
    \end{tabular}
\end{table}

\begin{table}[htb]
    \caption[Criteria Category “Conformance Checking”]{Criteria Category “Conformance Checking”}
    \label{tab:cc}
    \centering
    \begin{tabular}{>{\raggedright\arraybackslash}m{2cm}m{5.5cm}}
    \toprule
    \textbf{Criterion} & \textbf{Brief description}\\
    \midrule
   Compare As-Is and Target Process & Yes/No - User can compare as-is process with a target process, e.g. happy path\\
    \midrule
   Target Model Creation & Import model \ding{51}\ding{55} ($<$model types$>$), Auto-create from as-is \ding{51}\ding{55},
Create new \ding{51}\ding{55}
\\
    \midrule
   In-Graph Conformance Visualisation & Yes/No - Deviations from a target process can be visualised in the process graph\\
    \midrule
   List of Confor-\newline mance Violations & Yes/No - List of identified conformance violations for undesired activities, missing activities and non-compliant activity sequence\\
    \midrule
   Four-Eyes Principle & Yes/No - Breach of the four-eyes principle can be detected for any 2 selected activities\\
    \midrule
   Sequence Filtering & Yes/No -  User can filter by the condition “activity A is (not) directly followed by activity B”\\
    \midrule
   Conformance Root Cause Analysis & Yes/No - Root causes can be automatically identified for selected conformance violations\\
    \bottomrule
    \end{tabular}
\end{table}

\begin{table}[h!]
    \caption[Criteria Category “Operational Support”]{Criteria Category “Operational Support”}
    \label{tab:operational}
    \centering
    \begin{tabular}{>{\raggedright\arraybackslash}m{2cm}m{5.5cm}}
    \toprule
    \textbf{Criterion} & \textbf{Brief description}\\
    \midrule
   Alert Generation & Yes/No - Capability to trigger alerts defined by the user via query/filter, KPI threshold or a particular time interval\\
    \midrule
   Predictive Analytics & Yes/No - Capability to predict the future outcome of a running case based on historic data \cite{van2016}\\
    \midrule
   Recommendations (Prescriptive Analytics) & Yes/No - Capability to suggest potential next actions in order to meet a particular business goal, e.g. minimising cycle time \cite{van2016}\\
    \bottomrule
    \end{tabular}
\end{table}

\begin{table}[htb]
    \caption[Criteria Category “Views, Monitoring and Reporting”]{Criteria Category “Views, Monitoring and Reporting”}
    \label{tab:views}
    \centering
    \begin{tabular}{>{\raggedright\arraybackslash}m{2cm}m{5.5cm}}
    \toprule
    \textbf{Criterion} & \textbf{Brief description}\\
    \midrule
   Export Reports & Events ($<$formats$>$), Cases ($<$formats$>$), \newline Variants ($<$formats$>$)\\
    \midrule
   Export Charts and Tables & Yes/No\\
    \midrule
   Custom Dashboards & Custom charts \ding{51}\ding{55}, Custom tables \ding{51}\ding{55}\\
    \midrule
   Custom Metrics/ KPIs & Yes/No - User can define custom metric/KPI through a formula using own syntax, or by selection of any imported numerical attribute with the option of at least 5 different aggregation types, e.g. mean, median and percentiles\\
    \midrule
   KPI Thresholds & Yes/No - User can define thresholds for metrics/KPIs or charts to emphasise (non-)acceptable values by colour highlighting\\
    \midrule
   Advanced Charts & Yes/No - User can choose from at least 5 different chart types\\
    \midrule
   World Map & Latitude \& longitude coordinates \ding{51}\ding{55}, Location by attribute (e.g. country codes, city names) \ding{51}\ding{55} \newline (Visualisation of process-related locations in a world map graph)\\
    \midrule
   Save Filter Settings & Yes/No - Applied filter settings can be reused at a later point in time\\
    \midrule
   UI Languages & List of all available languages in the GUI\\
    \midrule
   Share and Collaborate & Share selection \ding{51}\ding{55}, $<$collaboration features$>$ (Sharing applied filter settings with other users; List of all additional means to collaborate and share insights, e.g. comment feature)\\
    \bottomrule
    \end{tabular}
\end{table}

\begin{table}[h!]
    \caption[Criteria Category “Advanced Enhancement Capabilities”]{Criteria Category “Advanced Enhancement Capabilities”}
    \label{tab:enhancement}
    \centering
    \begin{tabular}{>{\raggedright\arraybackslash}m{2cm}m{5.5cm}}
    \toprule
    \textbf{Criterion} & \textbf{Brief description}\\
    \midrule
   Organisational Mining & Yes/No - Capability to visually add organisational perspective by grouping of activities and org. entities such as resources and departments \cite{van2016}\\
    \midrule
   Scenario Simulation & Yes/No - The impact of specific process alternations (e.g. adjusting resource allocation and work times for activities) on the overall process can be simulated\\
    \midrule
   Decision Rule Mining & Yes/No - Automatic derivation of rules for decision points based on case-related data such as case-level attributes \cite{decisionrules}\\
    \bottomrule
    \end{tabular}
\end{table}

\begin{table}[htb]
    \caption[Criteria Category “Security and Compliance”]{Criteria Category “Security \& Compliance”}
    \label{tab:security}
    \centering
    \begin{tabular}{>{\raggedright\arraybackslash}m{2cm}m{5.5cm}}
    \toprule
    \textbf{Criterion} & \textbf{Brief description}\\
    \midrule
   Role-Based Access & Yes/No - Access to projects, dashboards or certain process data can be restricted for any user in the system via user roles or user-specific access permissions\\
    \midrule
   User Authentication & List of all means of authentication for user login, e.g. 2FA, LDAP, Active Directory\\
    \midrule
   Audit Logs & Yes/No - Audit logs can be produced which contain data of at least user identification, executed activity and corresponding timestamp\\
    \bottomrule
    \end{tabular}
\end{table}

\end{document}